\documentclass{chjaa}                  % preprint, the final version for publication
\usepackage{graphicx,times}             %for PS/EPS graphics inclusion, new
\input{epsf.sty}                        %for PS/EPS graphics inclusion, old
\input{psfig.sty}                       %for PS/EPS graphics inclusion, old

\begin{document}

   \title{X-ray plateaus followed by sharp drops in GRBs 060413, 060522, 060607A and 080330:
          Further evidences for central engine afterglow from Gamma-ray Bursts}

%  The X-ray plateau followed by a sharp drop in GRB 060607A: Another case
%    of the power-law decaying central engine afterglow

   \volnopage{Vol.0 (200x) No.0, 000--000}      %%preserved for Editor. DOn't remove!
   \setcounter{page}{1}           %%starting page, preserved for Editor. DOn't remove!

   \author{X. H. Zhang }
      \inst{1,3}\mailto{}
%% Please move "\mailto{}" to the corresponding author of the paper.

   \institute{
$^1$  Department of Physics, Yunnan University, Kunming 650091, China\\
             \email { xhzhang2008@yahoo.cn
             }}

%% offprint requests will be sent. Note to pair \mailto{} with \email{}

   \date{Received~~2001 month day; accepted~~2001~~month day}

   \abstract{
\baselineskip=18pt The X-ray afterglows of  GRBs 060413, 060522,
060607A and 080330 are characterized by plateaus that are followed
by very sharp drops. An X-ray plateau is interpretable within the
framework of the external forward shock model but the sharp drop is
not. In this work we interpret these peculiar X-ray afterglow data
as the central engine afterglows from some magnetized central
engines, plausibly magnetars. In this model, the X-ray afterglows
are powered by the internal magnetic energy dissipation and the
sudden drop is caused by the collapse of the magnetar. Accordingly,
the X-ray plateau photons should have a high linear polarization,
which can be tested by the future X-ray polarimetry.
 }

  \authorrunning{}            %author_head in even pages
   \titlerunning{Further evidences for central engine afterglow from Gamma-ray Bursts }  % title_head in odd pages

\maketitle
%% The author head (on even pages) and the title head (on odd pages) will be
%% automatically extracted from \author{} and \title{}. Whenever the title is too long,
%% you will be asked to supply a shorter one by inserting either \authorrunning{} or
%% \titlerunning{} before \maketitle. Anyway, you can specify your own heads in advance.
%%
%%
%% Note: In the following text body of your manuscript, please note several differences from
%%       other major journals:
%% (1) \subsection{Please Capitalize the First Letter of Each Notional Word in Subsection Title}
%% (2) Please Capitalize the First Letter of Each Notional Word in all tables' captions

%
%________________________________________________ sections below
%
\section{Introduction}           %% first-level sections will be auto-capitalized
\label{sect:intro}

Gamma-Ray Bursts (GRBs) are by far the most luminous objects in the
universe. They are bright flashes of high energy photons. A typical
GRB usually lasts about several or tens of seconds. In the standard
fireball model, the prompt $\gamma$-ray emission is powered by the
internal shocks (Paczynski \& Xu 1994; Rees \& M\'{e}sz\'{a}ros
1994; Kobayashi et al. 1997; Daigne \& Mochkovitch 1998; Spruit et
al. 2001; Fan, Wei \& Zhang 2004), and the GRB afterglows are the
emission of the external forward shock driven by the GRB fireball
expanding into the surrounding medium (M\'{e}sz\'{a}ros \& Rees
1997; Sari et al. 1998; Dai \& Lu 1998a; Chevalier \& Li 2000; Huang
et al. 2000). The fireball afterglow model has been widely accepted
because it works pretty well on reproducing the late time
multi-wavelength afterglow data in the pre-{\it Swift} era (e.g.,
Sari et al. 1998; Panaitescu \& Kumar 2002; Piran 2004).

The GRB central engine may also play an important role in producing
afterglow emission, i.e., the so-called ``central engine afterglow"
(Katz, Piran \& Sari 1998; Fan, Piran \& Xu 2006). One disadvantage
of this model is its lack of predictive power. Though somewhat ad
hoc, people began to interpret the data with this model. For
example, Piro et al. (1998) discovered a late-time outburst of the
X-ray afterglow of GRB 970508 and attributed such an outburst to the
re-activity of the central engine. However, the energy injection
model can reproduce the multi-wavelength outburst data quite well
(\cite{PMR98}). So the late time outburst detected in GRB 970508 is
not a good {\it central engine afterglow} candidate. The situation
changed dramatically in 2005.  Piro et al. (2005) discovered two
very early X-ray flares in GRB 011121, which had been interpreted as
the central engine afterglow---the prompt emission powered by the
re-activity of the central engine (Fan \& Wei 2005). Since then,
more and more X-ray flares have been well detected in early
afterglow (e.g., Nousek et al. 2006), and their central engine
origin has been well established (e.g., Zhang et al. 2006).

In view of the similarity of the temporal behaviors of X-ray flares
and GRBs, it is not a surprise to see that they have a common
origin. Central engine afterglows may be a bit more common than
previously thought. People found out that even some power-law
decaying X-ray afterglows might have a central engine origin. A good
example may be the afterglow of GRB 060218. As shown in Fan et al.
(2006), the inconsistence of the X-ray afterglow flux with the radio
afterglow flux and the very steep XRT spectra support here the
central engine afterglow hypothesis (see \cite{ZLZ07,Liang2008} for
more cases). Recently, Troja et al. (2007) argued that the X-ray
plateau followed by a sharp drop in GRB 070110 is also a central
engine afterglow.

In this work, we discuss the optical and X-ray afterglow of GRB
060607A. We show that not only the early X-ray plateau followed by a
sharp drop but also the very early optical re-brightening may be due
to the prompt emission of the long activity of the central engine.
This result suggests that the central engine optical/X-ray
afterglows might be common. We also apply the central engine
afterglow model to quite a few more bursts, including GRB 060413,
GRB 060522 and GRB 080330.

%% Authors can use \cite, \citep and \citet for citation.
%% You may also give a citation as 'Michel et al. 1992', and use Table~1 or Fig.~1
%% and so forth. Using \ref and \label for cross-references of Tables/Figures is
%% a good way in adjusting/adding/removing text, tables or figures.

\section{Observations}
\label{sect:Obs} 
\begin{figure}
\centering
\includegraphics[width=0.7\textwidth]{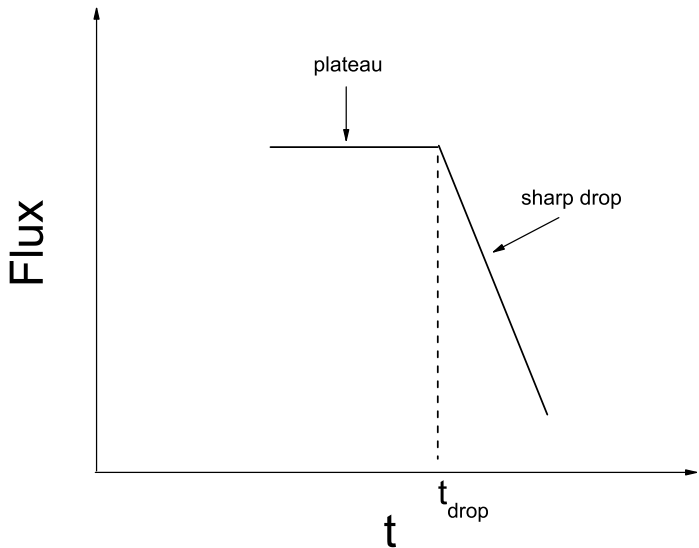}
\caption{---The plateau followed by a sharp drop is another central
engine afterglow.  }\label{f4}%{\bf [The figure should be revised. It
                              %should be "Log (t)" and "Log (X-ray Flux)"]}
\end{figure}

GRBs 060413, 060522, 060607A, 070110 and 080330 are
different in many aspects, for example, their redshifts and the
isotropic energies of the prompt $\gamma-$ray emission
(\cite{Moli07, Liang2007, Tueller 2006, Ledoux2006}) . However,
their X-ray afterglows share a common character. As schematically
plotted in Fig.\ref{f4}, an X-ray flat segment with a luminosity
$L_{_{\rm X}}\sim 10^{47}-10^{48}~{ erg~s^{-1}}$ is evident in the
early afterglow phase (see Tab.1 for details). The plateau
disappeared suddenly with a decline as steep as $t^{-4}$ or even
$t^{-8}$. The optical afterglow of GRB 060607A is characterized by
an optical flare hidden in the first optical re-brightening. The
optical flare drops with time as $t^{-11}$ (see Figure 2 of
Nysewander et al. 2007). These peculiar behaviors are inconsistent
with the regular afterglow model and may shed some lights on the
central engine.

\section{Interpretation of the data}
\label{sect:data} Quite a few physical processes are able to give
rise to an X-ray flat segment. Below we discuss three of them,
including the energy injection model, the density jump model, and
the central engine afterglow model.

\subsection{Energy injection model} We assume that the GRB central engine
does not die after the prompt $\gamma-$ray emission. Such a long
living central engine has an energy output $L \propto t^{-q}$ (Cohen
\& Piran 1999 and Zhang \& M\'esz\'aros 2001), where $q \geq 1$
represents the weak energy injection and $q = 0$ corresponds to the
early time energy injection from a fast rotating pulsar/magnetar
(Dai \& Lu 1998b; Zhang \& M\'esz\'aros 2001; Dai 2004; Fan \& Xu
2006). Other $q$ values are possible if an energy injection results
from slower material progressively catching up (viz. the Lorentz
factor of a GRB has a wide range) (Rees \& M\'esz\'aros 1998; Kumar
\& Piran 2000) or if an energy injection is caused by the fall-back
of the envelope of the massive star (MacFadyen, Woosley \& Heger
2001).

If the energy injection is significant enough, the dynamics of the
forward shock will be modified. So is the afterglow light curves
(\cite{zhang2006}). For the X-ray plateau, $\nu
> \max \{\nu_c, \nu_m\}$ is usually satisfied, where $\nu$ is the
observer's frequency, $\nu_c$ is the cooling frequency and $\nu_m$
is the forward shock typical synchrotron radiation frequency (Sari
et al. 1998). Under this condition, the temporal decline index of
the X-ray afterglow should be
\begin{equation}
\alpha={(2p - 4) + ( p+ 2)q \over 4},
\end{equation}
where $p \sim 2.3$ is the energy distribution power-law index of the
shocked electrons in the blast wave. In GRB 060607A, the decay of
the plateau before the sharp drop can be approximated $t^{-0.1}$.
Taking $p\sim 2.3$, we have (note that now $\alpha \sim 0.1$)
\[
q\sim 0,
\]
which is consistent with a pulsar/magnetar energy injection. Similar
conclusions can be drawn for X-ray flat segments detected in GRBs
060403, 060522, 070110 and 080330, too.

In this model, the X-rays are the synchrotron radiation of electrons
accelerated by the forward shock. Their decline is determined by the
dynamics of the forward shock and by the spherical curvature of the
blast wave. Usually the decline should be shallower than $t^{-p}$
(see Piran 2004 and Zhang 2007 for reviews). With a typical $p\sim
2.3$, the decline shallower than $t^{-2.3}$ is not steep enough to
account for the sharp drops detected in GRBs 060413, 060522,
060607A, 070110 and 080330 ($\alpha=4\sim8$). So the energy
injection model is disfavored in these particular cases.

\subsection{Density jump model} Dai \& Lu (2002) proposed a density
jump model to account for some afterglow re-brightening. In this
model, if the density jump is large enough, strong reverse shock
emission forms and will give rise to strong X-ray/UV/optical
emission (cf. Nakar \& Granot 2007). The decline of the afterglow
light curves will be suppressed, too. So it is potentially to
account for the X-ray flat segments that detected in GRBs 060413,
060522, 060607A, 070110 and 080330. However, currently the forward
and reverse shock regions move with the same bulk Lorentz factor.
The curvature effect of the reverse shock emission is the same as
that of the forward shock emission. As shown in many literatures
(e.g., Fenimore et al. 1996; Kumar \& Panaitescu 2000; Fan \& Wei
2005; Zhang et al. 2006; Liang et al. 2006), because of the
spherical curvature of the emitting region and because of the
special relativity effect, the external forward shock emission can
not drop with time more quickly than
\begin{equation}
F_\nu \propto t^{-(2+\beta)},
\end{equation}
where $\beta$ is the spectral index of the X-ray emission.
 The only
exception is that the ejecta is so narrow that we have seen its
edge. If true the optical emission would show the same behavior,
which is not the case. We then conclude that the density jump model
is not a good candidate to interpret the X-ray plateaus followed by
sharp drops.

Recently Nysewander et al. (2007) employed the density jump model to
account for the optical re-brightening detected in GRB 060607A. We,
however, find out that the first optical flare hidden in the
re-brightening (see their Figure 2) has a very steep decline $\sim
t^{-11}$, which violates the curvature effect constraint seriously.
Therefore, such an interpretation is problematic.

\subsection{Central engine afterglow model}
Some authors have discussed the energy injection of a
pulsar/magnetar wind and its influence on the afterglow emission
(Dai \& Lu 1998b and Zhang \& M\'esz\'aros 2001). Strong magnetic
energy dissipation (e.g. via reconnection) may take place if the
continued outflow from the central engine is Poynting-flux dominated
(Fan, Zhang \& Proga 2005; Gao \& Fan 2006; Giannios 2006), as that
happened in the phase of prompt $\gamma-$ray emission (Usov 1994;
Thompson 1994; Lyutikov \& Blandford 2003). In this case, part of
the magnetic energy of the outflow will be converted into the
delayed prompt emission, maybe mainly in X-ray band, before they
inject into the forward shock.

Here following Gao \& Fan (2006) and Troja et al. (2007), we discuss
the central engine afterglow powered by a millisecond magnetar. It
is well known that the spin-down timescale of a millisecond magnetar
can be estimated as
\begin{equation}\label{T_em}
T_{\rm em} = 4 \times 10^{3}~{\rm s}~(1+z)
I_{45.3}B^{-2}_{p,15}P^{2}_{0,-3}R_{\rm s,6}^{-6},
\end{equation}
where $B_{p,15}= B_{p} / 10^{15}$ Gauss is the dipole radiation
magnetic field of the pulsar in units of $10^{15}$ Gauss, $P_{0,-3}$
is the initial rotation period in milliseconds, $I_{45.3}$ is moment
of inertia in units of $2\times 10^{45} ~{\rm g~cm^{2}}$, $R_{\rm
s,6}$ is the stellar radius in units of $10^{6}$ cm.

The dipole radiation luminosity is
\begin{eqnarray}\label{L}
L_{\rm dip} \approx 10^{49}~{\rm erg~s^{-1}}~B_{\rm p,15}^2R_{6}^6
P_{0,-3}^{-4}[1+{t\over T_{\rm em}}]^{-2}. \label{eq:E_inj}
\end{eqnarray}
We have $L_{\rm dip}\propto t^0$ for $t\ll T_{\rm em}$ and $L_{\rm
dip}\propto t^{-2}$ for $t\gg T_{\rm em}$.

For such a Poynting-flux-dominated flow, the dissipation of the
magnetic fields may produce X-ray/$\gamma$-ray emission (Usov 1994;
Zhang \& M\'esz\'aros 2002; Fan et al. 2005). In this work we adopt
the so-called MHD approximation breakdown model (Usov 1994). By
comparing with the pair density ($\propto r^{-2}$, $r$ is the radial
distance from the central source) and the density required for
co-rotation ($\propto r^{-1}$ beyond the light cylinder of the
compact object), one can estimate the radius $r_{\rm MHD}$ at which
the MHD condition breaks down $r_{\rm MHD} \sim (2\times 10^{15})
L_{\rm dip, 48}^{1/2} \sigma_1^{-1}t_{v,m,-3} \Gamma_2^{-1} ~{\rm
cm}$ (e.g. Zhang \& M\'esz\'aros 2002; Fan et al. 2005), where
$\sigma$ is the ratio of the magnetic energy flux to the particle
energy flux, $t_{v,m}$ is the minimum variability timescale of the
central engine, and $\Gamma$ is the bulk Lorentz factor of the
outflow. Beyond this radius, intense electromagnetic waves are
generated and outflowing particles are accelerated  (Usov 1994).
This converts magnetic energy into radiation. At $r_{\rm MHD}$, the
comoving magnetic fields $B_{\rm MHD}$ can be estimated as  $B_{\rm
MHD}\sim 50~\sigma_1 t_{v,m,-3}^{-1}~{\rm Gauss}$. At $r_{\rm MHD}$,
the typical synchrotron radiation frequency of the
accelerated-electrons can be estimated as (Fan et al. 2005)
\begin{equation}
\nu_{\rm m,MHD}\sim6\times10^{16}\sigma_1^{3}C_{p}^{2} \Gamma_{2}
{t_{v,m}}_{-3}(1+z)^{-1}~{\rm Hz},
\end{equation}
where
$C_{p}\equiv(\frac{\epsilon_{e}}{0.5})[\frac{13(p-2)}{3(p-1)}]$,
$\epsilon_{e}$ is the fraction of the dissipated comoving magnetic
field energy converted to the comoving kinetic energy of the
electrons, and the accelerated electrons distribute as a single
power-law $dn/d\gamma_{e}\propto\gamma_{e}^{-p}$. According to this
model, a $\sigma$ much larger than $10$ is disfavored.

The cooling Lorentz factor of the accelerated-electrons can be
generally estimated as $\gamma_{\rm e,c}\sim 4.5\times 10^{19}\Gamma
/(r_{\rm MHD} B^2)$, which is $\sim 10^{3}$ for typical parameters
taken here and is comparable to $\gamma_{\rm e,m}$. So the
synchrotron radiation of the accelerated electrons peaks in the soft
X-ray band. The energy emitted in X-ray band, of course, is just a
fraction ($\epsilon_{\rm x}$) of the total magnetic energy
dissipated, so we have
\begin{equation}
L_{_{\rm X}} \sim \epsilon_{\rm x}L_{\rm dip} \sim 10^{48}~{\rm
erg~s^{-1}}~\epsilon_{\rm x,-1} B_{\rm p,15}^2R_{6}^6
P_{0,-3}^{-4}[1+{t\over T_{\rm em}}]^{-2}. \label{eq:E_inj}
\end{equation}
Now we have a set of free parameters $({B_{\rm p}, P_{0}}, I, R_{\rm
s}, \epsilon_{\rm x})$ but just two observation data, $L_{_{\rm X}}$
and the sharp drop time
\begin{equation}
t_{\rm drop} \sim f T_{\rm em}\sim 4 \times 10^{3}~{\rm s}~ (1+z)f
I_{45.3}B^{-2}_{p,15}P^{2}_{0,-3}R_{\rm s,6}^{-6},
\end{equation}
where $f$ is also a free parameter. So these free parameters can be
fully determined by the limited observation data. Below we fix
$(B_{\rm p}, P_{0}, I, R_{\rm s})$ with the typical values $\sim
(10^{15}~{\rm Gauss},1{\rm ms},2\times 10^{45}{\rm g~cm^2},10^6{\rm
cm})$ and then constrain $\epsilon_{\rm x}$ and $f$, respectively.
The results are shown in Tab.\ref{tab:1}. We do not want to
over-interpret the results because the typical values of $(B_{\rm
p}, P_{0}, I, R_{\rm s})$ adopted here might be biased, so are the
derived $\epsilon_{\rm x}$ and $f$. Anyhow for GRB 060413, GRB
060607A and GRB 070110, the parameters are largely reasonable and
thus support the millisecond magnetar central engine hypothesis. For
GRB 060522 and GRB 080330, $E_{_{\rm X}} \approx t_{\rm
drop}L_{_{\rm X}}/(1+z) \sim 10^{50}$ erg  is significantly smaller
than the total rotation energy of a millisecond magnetar $\sim
I(2\pi/P_0)^2/2 \sim 4\times 10^{52}I_{45.3}P_{0,-3}^{-2}$ erg.
Possibly these two magnetars were hypermassive and collapsed before
losing significant fraction of the rotation energy and angular
momentum (Fan \& Xu 2006). Alternatively the magnetic field of the
magnetars decayed suddenly for some unknown reasons (Troja et al.
2007). The current data are not sufficient for us to distinguish
between these two possibilities. In the future the gravitation wave
observation may shed some lights on the nature of the X-ray drop
because the collapse of the hypermassive magnetar may give rise to
interesting gradational signals.

 \begin{table}[!htbp]
%      \caption{\label{tab:1}Several GRBs central engine afterglow for X-ray band and optical band(V)}
%        \label{tab:1}
        \caption{\label{tab:1}}

        \centering
        \footnotesize

         Fit to the central engine afterglows of several GRBs, where $B_p$,
         $R_s$, $I$
         and $P$ are fixed.\\
        \begin{tabular}{rrrrrrr}

        \hline\hline
         GRBs    & $t_{\rm drop}$(s) &  $z$      &$L_{_{\rm x}}(erg/s)$& $\epsilon_{\rm x}$ & $f$ & $F_{\nu_{\rm v}}/F_{\nu_{\rm x}}$\\
        \hline
        GRB 060413& $2\times10^{4}$ &  3      &$1.02\times10^{48}$    &      0.4          &  1.2   &      3.4        \\

        GRB 060522  & $4\times10^{2}$& 5.11   & $2.48\times10^{48}$&         0.2          &   0.02  &        5.2           \\

        GRB 060607A & $1.5\times10^{4}$& 3.08  &$1.71\times10^{48}$&         0.4           &  0.9  &        10        \\

        GRB 070110  & $2\times10^{4}$  & 2.35  &$2.7\times10^{47}$ &         0.1          &  1.4  &        1.4      \\

        GRB 080330  & $1.0\times10^{3}$ & 1.51 &$1.12\times10^{47}$ &        0.01           &   0.1 &         62      \\

        \hline

        \end{tabular}

        Reference.----http://www.swift.ac.uk/, \cite{Liang2007, Troja2007}.\\
        Notes.----- For GRB 060413 we set $z\sim 3$, the typical redshift of Swift GRBs.
\end{table}

The prolonged activity of the central engine should also produce
some emission in the optical band. A simple estimate is the
following. The synchrotron self-absorption frequency can be
estimated as (e.g., Fan \& Wei 2005)
\begin{equation}
\nu_{\rm a}\sim 2\times 10^{14}~{\rm Hz}~[2/(1+z)]^{3/7}L_{\rm dip,
48}^{2/7}\Gamma_{2}^{3/7}r_{\rm MHD,15}^{-4/7}B_{\rm MHD,2}^{1/7},
\end{equation}
which is below the optical band $\nu_{\rm v}\sim 5\times 10^{14}$
Hz. With $\gamma_{\rm e,m} \sim \gamma_{\rm e,c}$ and $\nu_{\rm m,
MHD} \sim \nu_{_{\rm X}}\sim 7.2\times 10^{16}$ Hz, the optical
central engine afterglow flux is not expected to be dimmer than
\begin{equation}
F_{\nu_{\rm v}}\sim F_{\nu_{_{\rm X}}}(\nu_{\rm v}/\nu_{_{\rm
X}})^{1/3}\sim 0.2 F_{\nu_{_{\rm X}}}.
\end{equation}
Consequently the observed optical emission should be brighter than
$\sim 0.2 F_{\nu_{_{\rm X}}}$, which is consistent with the
observations, as shown in Tab.\ref{tab:1}.\\

\section{Discussion and Conclusions}
\label{sect:conclusion}
 In contrast to what was believed in the
pre-{\it Swift} era, it is evident now that the central engine plays
an important role in producing afterglow emission (see Fan, Piran \&
Wei 2008 for a review). Most theoretical works so far focus on the
energetic flares that are well detected in many {\it Swift} GRBs.
The temporal behavior of the flaring X-rays are quite similar to
that of the prompt $\gamma-$rays. It is thus reasonable that the
flares and the prompt GRB have a common origin. In this work, we
show that the X-ray plateaus followed by sharp drops detected in
GRBs 060413, 060522, 060607A, 070110 and 080330 are also good
candidates of the so-called ``{\it central engine afterglow}" (see
also Jin \& Fan 2007; Troja et al. 2007; Staff et al. 2007).
%So is the optical flare hidden in the first optical re-brightening
%for GRB 060607A, if its decline is as steep as $t^{-11}$ (see Figure 2
%of Nysewander et al. 2007).
The energy injection model and the density jump model are less
favored. We also find out that both the luminosity and the timescale
of the X-ray plateaus detected in GRBs 060413, 060522 and 070110 are
well consistent with the central engine afterglow emission powered
by millisecond magnetars (see section 3.3 for details). For GRB
060522 and GRB 080330, the X-ray drop appeared so early that might
suggest that the two magnetars, if they were, had collapsed before
losing a significant part of their rotation energies and angular
momentums. Though millisecond magnetar are believed to be a natural
outcome of the collapse of a massive star, the identification of
such a kind of compact objects at cosmological distances is not
easy. Fairly speaking, additional independent signature, like a high
linear polarization (see Fan, Xu \& Wei 2008), is needed before the
magnetar wind dissipation model for the X-ray plateaus that are
followed by X-ray drops can be generally accepted.

\begin{acknowledgements}
I thank the anonymous referee for insightful comments, and Prof. L.
Zhang and Dr. Y. Z. Fan for their kind help.
\end{acknowledgements}

\appendix                  %%appendicial material is supported

\end{document}